\documentclass[twocolumn,showpacs,preprintnumbers,amsmath,amssymb]{revtex4}

\usepackage{graphicx}
\usepackage{dcolumn}
\usepackage{bm}

\usepackage{graphicx}
\usepackage{amssymb}
\usepackage{epstopdf}
\newcommand{\ignore}[1]{}
\newcommand{\be}{\begin{equation}} \newcommand{\ee}{\end{equation}}
\newcommand{\ba}{\begin{eqnarray}} \newcommand{\ea}{\end{eqnarray}}
\newcommand{\nn}{\nonumber} \renewcommand{\bf}{\textbf}
\newcommand{\ra}{\rightarrow}

\renewcommand{\a}{\alpha}

\def\slasha#1{\setbox0=\hbox{$#1$}#1\hskip-\wd0\hbox to\wd0{\hss\sl/\/\hss}}
\def\slashb#1{\setbox0=\hbox{$#1$}#1\hskip-\wd0\dimen0=5pt\advance
       \dimen0 by-\ht0\advance\dimen0 by\dp0\lower0.5\dimen0\hbox
         to\wd0{\hss\sl/\/\hss}}

\input{epsf}

\begin{document}

\title{Interferometric Parallax:  \\ A Method 
for Measurement of Astronomical Distances}

 \author{Pankaj Jain$^1$ and John P. Ralston$^2$}
 \affiliation{$^1$Physics Department, IIT, Kanpur 208016 \\ 
$^2$Department of Physics and Astronomy, University of Kansas, Lawrence KS 66045}
 \input{epsf}

\begin{abstract}
   We show that distances of objects at cosmological distances can be measured directly using interferometry. 
Our approach to interferometric parallax comes from analysis of 4-point 
amplitude and intensity correlations that can be generated from pairs of 
well-separated detectors. 
The baseline required to measure cosmological distances of Gigaparsec order 
are within the reach of the next generation of space-borne detectors. 
The semi-classical interpretation of intensity correlations uses a notion of a single photon taking two paths simultaneously. 
Semi-classically a single photon can simultaneously enter four detectors separated by an astronomical unit, developing correlations feasible to measure with current technology. 
\end{abstract}


\maketitle

 There is no more important problem in astronomy than resolving the third dimension of source distances. The crisis of dark energy and dark matter in cosmology hinges on distance measurements. Estimates using red shifts or Type 1a supernova sources are weakened by model dependence of the cosmology and assumptions on the evolution of distant sources.  Here we show that direct measurements of cosmologically distance objects can be made from the analysis of correlations of detectors separated on the scale of the solar system.   Correlations of amplitudes (first order coherence) are used in Michelson's interferometer and radio telescopes. The breakthrough of Hanbury-Brown and Twiss ($HBT$)\cite{HBT}  demonstrated 2-point correlations of intensity (second order coherence) developed by counting photon fluxes in separated detectors. Higher order correlations have been 
 proposed earlier for reconstructing the phase of the coherence function
\cite{phase} and to improve the sensitivity \cite{sensitivity} in intensity correlations. We show that
4-point amplitude and intensity correlations contain further information on the distance to the sources. The baseline required to measure cosmological distances of Gigaparsec order are within the reach of the next generation of space-borne detectors.  Measuring source distances of Megaparsec order appears feasible now.    
\begin{figure}
\includegraphics[width=3in,height=2.5in]{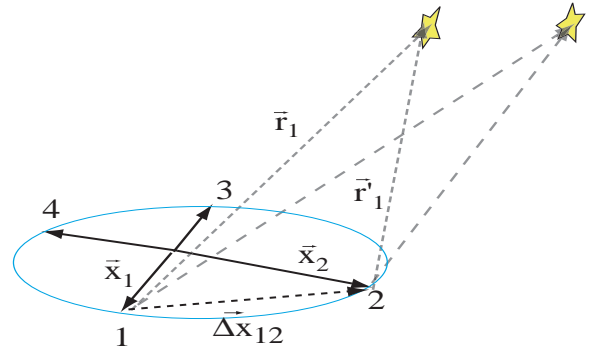}
\caption{Four detectors are positioned so that the relative position
vector $\Delta \vec x_{12}$ between one pair is approximately the same
as between another pair $\Delta \vec x_{34}$, with a net translation $\vec X$ between the pairs.   }
\label{fig:geometry}
\end{figure} 

Let $\vec x_{i} $ be the position vectors of the $i$th detector 
relative to the origin (Fig. \ref{fig:geometry}). Consider a point source at $\vec r = \hat r r$ relative 
to the origin, also located at position $\vec r_{J}$ relative to each detector:  \ba  \vec r_{J} = \vec r- \vec x_{J}. \nn \ea 
Calculate the distance from the $J$th detector to the 
source to order $1/r$: 
\ba r_{J} = r - \hat r \cdot \vec x_{J} +{ x_{J}^{i}\delta_{T}^{ij} x_{J}^{j} 
\over 2r} +O(1/r^{2}),\ea where $\delta_{T}^{ij}(r) = \delta^{ij} -\hat r^{i}
\hat r^{j}.$  
Here upper indices denote vector components.
The third term depends on the distance $r$ and will be responsible for probing it via interferometric parallax.  The frequency-domain Green function for propagation from source to receiver is expanded \ba  G_{x_{J} \vec r} &=&  { e^{i k |\vec r- \vec x_{'J}|} \over  |\vec r- 
\vec x_{J}|}  \sim  {e^{ i k r} \over r}  e^{ - i k \hat r \cdot 
\vec x_{J}} e^{ i k { x_{J}^{i}\delta_{T}^{ij} x_{J}^{j} /( 2r}) } \nn \ea 
where $k$ is the wave number.  We drop the $ e^{ i k r }/r$ prefactor which cancels out in calculations. The formula extends trivially to two distinct sources, $S$ and $S'$, 
for which primed symbols such as $\vec r' =\hat r' r'$ take the obvious meaning.

Concentrate for a moment on two detectors 1, 2.  Standard analysis of the interferometric correlation treats the electric field $\vec E$ as a random variable,  described by convolution of the Green function with correlations of the source. Then the amplitude correlation for a single polarization between the two 
receivers can be expressed as,
\be
<E_1E_2^*> = e^{-ik\psi} \Bigg[ {I_0\over r^2} + 
{I_0'\over r^{\prime 2}}  e^{-i\phi_{12}^{(0)}
-i\phi_{12}^{parallax}}\Bigg]
\label{eq:amplitude_corr}
\ee
where $\vec E_{0}$ is the field of one source, $I_0 = <E_0E_0^*>$, and so on for primed variables. Here \ba    \phi_{12}^{(0)} &=& -k (\vec x_{1}-\vec x_{2})\cdot
(\hat r- \hat r'); \nn \\ \phi_{12}^{parallax} 
&=&  k ( \vec x_{1}^{i} \vec x_{1}^{j}-\vec x_{2}^{i} \vec x_{2}^{j})
\left( { \delta_{T}(r)^{ij} \over 2 r} -{ \delta_{T}(r')^{ij} \over 2 r'} 
\right).  \label{result} \ea 
The overall phase $\psi = \hat r\cdot (\vec x_1 - \vec x_2) - 
(x_1^ix_1^j - x_2^i x_2^j)\delta_T^{ij}/2r$. The overall phase cancels
in a determination of the absolute value of the correlation.
In eq. \ref{eq:amplitude_corr}, the source-detector 
distances are consistently replaced by $r$ and $r'$ everywhere except in the phases.

The intensity-intensity correlations,
$<I_{1}I_{2}>= <|E(\vec x_{1})|^{2}  |E(\vec x_{2})|^{2}  > $ between receivers 1 and 2 for classical light are found to be, 
\ba <I_{1}I_{2}> &=& 
\left({I_0\over r^2}+{I'_0\over r'^2}\right)^2
+{I_0^2\over r^4} +{{I'}_0^2\over r^{'4} } \nn \\ &+&
{2I_0 I'_0} \Re( \, G_{1r}G^*_{1r'}G_{2r'}G_{2r}^{*}\,)
\nn \\ 
&=& \left({I_0\over r^2}+{I'_0\over r'^2}\right)^2+  {I_0^2\over r^4}+
{{I'}_0^2\over r^{'4} }+ \nn \\ 
&+&  {2I_0I'_0\over r^2r'^2}\Re\left( e^{ i\phi_{12}^{(0)}+i\phi_{12}^{parallax} }
\right). 
\label{C12}
\ea 
Here $\Re $ denotes the real part.
A quantum mechanical calculation gives a similar result, and incorporates photon bunching.  We see that $\phi^{parallax}$ appears in both amplitude 
and intensity correlations.  Intensity correlations can be used at optical frequencies by simply counting photons, and have certain advantages in automatically canceling the overall phase $\psi$.
 
The first phase $\phi^{(0)}$ has been used extensively for angular position
measurements. Notice that this phase is translationally invariant - it depends only on the relative position vector of the two detectors. The scale $\Delta x_{12}$ is conjugate to a {\it difference} of wave numbers, {\it not the radiation wavelength}. Our focus is on the second phase $\phi_{12}^{parallax}$. Consulting Fig. \ref{fig:geometry}, let detectors 1 and 2 be located at 
average position $\vec X$ and separated by $\Delta \vec x_{12}$: 
$ \vec x_{1} =\vec X-\Delta \vec x_{12}/2, \, 
\vec x_{2} =\vec X + \Delta \vec x_{12}/2 .$ 
By substitution the second phase is 
\ba \phi_{12}^{parallax}= -k \Delta \vec x_{12} \cdot \left( { \delta_{T}(r) 
\over r} -{  \delta_{T}(r') \over r' }\right )\cdot \vec X.  \ea  
This phase will 
change with $\vec X$ even if $\Delta \vec x_{12}$ is fixed. The 
explanation of course is parallax sensed via curved wave fronts.  A measurement of the correlation's dependence on translating a fixed detector pair in $\vec X$ can probe the distance dependence on $1/r$ and $1/r'$.

We emphasize that interferometric parallax is qualitatively different
from standard trigonometric parallax long used in astronomy.  In trigonometric parallax the distance is estimated by a precise measurement of the angular position of the source from two different locations.  For interferometric parallax a precise measurement of the angular position is
not required, and dependence on the angular position is weak.  Strictly speaking, the individual sources need not be resolved.  Sensitivity exists in selecting source for the measurement, and excluding background.

We now discuss finite source size effects. The basic two point amplitude correlation for an incoherent source component located at $\vec r_{i}+\vec y$ is 
\begin{eqnarray}
<E(\vec x_1) E^*(\vec x_2)> &=& \int d^2 y <E_0(\vec y) E_0^*(\vec y)>
\nonumber\\
&\times &
{e^{ik[|\vec r_1 + \vec y| - |\vec r_2 + \vec y|}\over |\vec r_1 + \vec y|
|\vec r_2 + \vec y| }, \nn\\ &=&  e^{ik(r_1-r_2)} {\cal I}.
\label{eq:finite1}
\end{eqnarray}
Here we have expanded the argument of the exponential integrand as
\ba 
|\vec r_1 + \vec y| - |\vec r_2 + \vec y| = r_1 - r_2 + 
\hat r_1\cdot \vec y - \hat r_2\cdot \vec y \nonumber \\ -  (\vec y \cdot \hat r_{1})^{2}/2r +(\vec y \cdot \hat r_{2})^{2}/2r. \nn 
\ea 
By inspection the last two terms only contribute at order $1/r^{2}$. Let $a$ be the effective transverse size of the source.  All terms involved in ${\cal I}$ 
are functions of the angular size of the source $a/r$, and ${\cal I}$ is negligibly small if $a/r>> 1/(k\Delta x)$.  As long as $\Delta x  \lesssim a/(k r)$ the corrections due to finite size can be absorbed into the overall factors $I_0$,
$I_0'$ etc. in Eq. \ref{eq:amplitude_corr}, and a source appears to be a point.  This reproduces the usual planar source criterion\cite{optics} that an observable signal requires baselines smaller than the coherence zone of each source.  All higher order correlations for two sources can be expressed as products of such two point amplitude correlations of 
individual sources, justifying the point source approximation.  

{\it {Orders of Magnitude:}} For wavelength $\lambda$  and typical source angular separation $\Delta \theta$ we have $\phi_{12}^{(0)} \sim { \Delta \theta/(\lambda/\Delta x_{12} )}$ and $\phi_{12}^{parallax} \sim { X/r /( \lambda/\Delta x_{12})}.$  We recognize $\lambda/\Delta x_{12}$ as the lower limit on angular resolution from optics.   Similarly, the parallax phase is of order one if the baseline of translation $X$ could be resolved by an instrument of aperture $\Delta x$ looking from distance $r$.  Using the lower limit of the single-source coherence $k\Delta x \lesssim r/a$ gives 
$\phi_{12}^{parallax} \sim ( X/a)(1-r/r').$  For $\phi_{12}^{parallax} \sim 1$
and comparable $r, \, r'$ the source-size should match the translational scale.  Of course the coherence zone criteria do not require literally small sources, but represent the existence of Fourier modes (structure) in the regime of size indicated.  Setting $\phi_{12}^{parallax}\sim 1$ yields the distance scale that can be observed:  \ba r \lesssim 1 \, Gpc \, {X \over AU}{ \Delta x  \over AU}{ 1 {\rm mm} 
\over \lambda}. \ea 

Although phases can often be measured with exquisite accuracy, we will continue assuming $\phi^{parallax} \sim 1$ for our estimates. Consider detectors separated by $10^{4}$ km, a near Earth orbit, and translating over $X \sim 1 AU$ in a period of a year. Numerically 
\ba \phi_{12}^{(0)} \sim 10^{5} { \Delta \theta  \over arcsec}{{\rm 1 mm} 
\over  \lambda} {\Delta x_{12} \over 10^{4} km }; \nn \\   
\phi_{12}^{parallax} \sim 10^{-4} \,  { X \over AU} { 1 {\rm Gpc} \over r}
{  {\rm 1 mm} \over  \lambda} {\Delta x_{12}  \over 10^{4} km}.  \ea  
Baselines of order $10^4$ km at $cm$ wavelengths have been demonstrated by
current technology.  For Gpc distances, one needs to measure a 
relatively small phase or push the limits of baselines to much larger than
$10^4$ Km and/or wavelength to the sub-mm range. This may be possible due to
the huge range of possibilities for $\Delta \theta$. Quasar sources are 
believed to have physical sizes extending to the range of 1 AU, whereby $\Delta \theta \sim 10^{-9}\, arcsec$ at distances of $Gpc$ order.  The maximum 
coherence zone for such sources $\Delta x \sim \lambda r/a \sim 10^{11}m \, ( \lambda /mm)(r/Gpc)(AU/a)$ are compatible with baselines of order AU.  Black hole and GRB sources are of course even smaller, with correspondingly larger coherence zones.  It might also be possible to measure gravitationally lensed single objects, exploiting two path lengths $r$, $r'$.  
In principle measurement of $ \phi_{12}^{parallax}$ of order unity can measure distances to Gpc order provided suitable sources can be exploited. 

The ratio $\phi^{(0)}/\phi^{parallax}\sim 10^{8}$ for a typical value of 
$\Delta\theta \sim 0.1 arcsec$ assuming Gpc distance. There are reasons to expect high control over $\phi^{(0)}$ by technological means.  However for the rest of the paper we consider ``worst case'' scenarios, in which control of $\phi^{(0)}$ is less than ideal.  There happens to be a practical strategy to null out the effects of the rapidly varying phase. To control the effects 
of $\phi^{(0)}$ we can (in effect) measure it twice, using another pair of detectors 3, 
4, separated by the ``same'' offset: $ \Delta \vec x_{34} =\Delta \vec x_{12}  + \vec \eta\ . $ It is clear that $\vec \eta << \Delta \vec x_{12}$ can be made relatively small with great precision.  We
consider the product of $<I_1I_{2}><I_{3}I_{4}>$, which is one of the terms 
in the 4-point intensity correlation.  Denoting the overall normalizations by $N_1$
and $N_2$ we write 
\ba <I_1I_{2}> = N_1 + N_2 \cos(A+B); \nn \\<I_{3}I_{4}> = N'_1 + N'_2 \cos(A'+B'), \nn \ea 
where $A=\phi_{12}^{(0)}$, $B = \phi_{12}^{parallax}$, with the primes 
switching 12 $\ra 34$. In an experiment where $A >>B$ and $\Delta x_{12}\cdot 
(\hat r-\hat r')$ varies rapidly the products with an odd 
number of cosines will average to zero. The average here (symbol $<<\, >>$) 
might occur over running time in which drifts of 
the detector position values cause $\phi^{(0)}$ to vary.  Another cause 
of variation lies in small ranges in $\omega$ differing between
the detectors, and there are no doubt other possible causes. The term with 
two cosines gives 
\ba <<  \, \cos(A+B)\cos(A'+B')\, >> =   \nn \\  <<  \, \cos A \cos A'  \, >>\cos B \cos B'  \nn \\ +<<  \, \sin A  \sin A'  \, >>\sin B \sin B' . \nn  \ea  
The only non-zero data will come from $<<  \, \cos A \cos A'  \, >> = <<  \, 
\sin A \sin A'  \, >> =1/2$, namely those regimes when the rapidly varying 
terms coincide. 

Collecting the terms gives  
\ba <I_1I_{2}><I_{3}I_{4}>=  N_1N'_1+ N_2N'_2 \cos(B-B') \nn \\ 
=  N_1 N_1'+ N_2 N_2'\cos(\phi_{12}^{parallax}-\phi_{34}^{parallax}). 
\label{eq:four_correlation}
\ea 
Since the net translation of the 3-4 receiver pair is independent of the 
12 pair, it is straightforward to arrange for the surviving slow oscillation 
to produce a net signal. A simple configuration (Fig. \ref{fig:geometry}) puts 
$ \Delta x_{12}=\Delta x_{34} = \Delta x, \, \vec x_{3}=-\vec X -\Delta x /2, \, \vec x_{4}=-\vec X +\Delta x /2 . $ The difference term $\vec \eta$ is relatively negligible and was 
dropped.  The parallax terms add, producing an oscillation going like 
$\cos(k \Delta x \cdot \vec X(1/r_{1} - 1/r_{2}))$. Thus with a near and 
a far source, where $1/r_{1} - 1/r_{2} \sim 1/r_{1}$, one can measure the 
distance to the sources by interferometric parallax.  Continuing with one 
source after another the entire Universe could be mapped out with a new ``cosmic distance ladder.''

So far we have presented two pairs of receivers and signal development via ``off-line'' correlation calculations.  There may be advantages to directly correlating the signal from four receivers, either at the amplitude or the intensity level.   Let $<I_1I_2I_3I_4>$ be the raw four-point intensity correlation.  Dropping terms that oscillate rapidly, and with $\vec \eta \ra 0$, a calculation gives
\begin{equation}
<I_1I_2I_3I_4> = {\cal N}_1 + {\cal N}_2
\cos (\phi_{12}^{parallax}-\phi_{34}^{parallax}).
\label{eq:fourpoint}
\end{equation}
Here ${\cal N}_1$ and ${\cal N}_2$ are normalizations 
depending on the intensity of the two
sources.  Note all the other terms in the four point 
correlation either vanish after statistical averaging or reduce to the two terms given in Eq. \ref{eq:fourpoint}.  The remarkable cancellations in the 4-point intensity correlation show that distances can be measured in terms of a standard statistical description of raw data. 

Similarly, the 4-point amplitude correlation (observable at radio frequencies) consists of sums of terms with phases $\phi^{(0)}$ and $\phi^{parallax}$, 
along with ``sum-phases'' of the form $(\vec k_{1}+ \vec k_{2})\cdot \Delta \vec x$. Sum-phases in conventional long-baseline interferometry\cite{optics,astron} tend to cause difficulty due to atmospheric fluctuation effects.  Intensity interferometry is much less sensitive, as positioning accuracy is set by the coherence time and not the wavelength \cite{utah}.  
Nevertheless both amplitude and intensity correlations contain $\phi^{parallax}$ - Eqs. \ref{result}, \ref{C12} are essentially a generalization of the VanCittert-Zernicke theorem\cite{optics} - and both can perform interferometric parallax measurements.  It would be premature to assess the technological advantages of either scheme here.  

Earlier we mentioned that exacting measurement of source positions is not required.  High precision of angular positions is now replaced by high precision of the two baselines $\vec \Delta x_{12}$ and $\vec \Delta x_{34}$. How demanding are the tolerances, and what can be done to ameliorate them?   An error of one part in $10^{8}$ represents a tolerable ranging error of order $km$ on the
$\Delta x \sim 1 AU$ baseline.  Proven satellite ranging techniques \cite{ranging} achieve accuracies superior by orders of magnitude.   We note in addition that $\phi^{(0)}$ might be predicted in advance by conventional means and removed by phase-shifting, heterodyning and signal-processing strategies, either``on-line" and  ``off-line".   Moreover, $\phi^{(0)}$ can be observed during running as a useful cross-check, and being constant under translation, can serve as a sort of absolute positioning standard. Another question is whether exacting timing resolution is needed to ensure the ``same photon'' enters all detectors.  The answer lies in the uncertainty principle.  Let $\Delta \omega$ be the bandwidth of detection, under which there is a variation
$\Delta \phi^{(0)} = \Delta \omega \phi^{(0)}/\omega$.   Suppose this error must be of order $10^{-8}$ or smaller, which is the relative phase error from position errors.  Then if position and bandwidth errors are in tolerance, the ``same wave'' enters the two detectors in tolerably fixed phase relation.  These criteria in fact define ``same wave'' and ``same photon'' concepts operationally. At optical frequencies $\Delta \omega /\omega \sim 10^{-8}$ allows errors of order $10 MHz$, integration over times of order $1 -10 \mu $s, which is no barrier.

The biggest unknowns appear to be technological and beyond the scope of 
this paper. Radio astronomy is so sophisticated in handling noise and 
phase variables that detailed estimates must be left to experts.  For 
optical intensity correlations we can nevertheless investigate whether 
there might be fundamental limitations from photon statistics. From the 
uncertainty principle there is a relation $\Delta n \Delta 
\phi>1$, where $\Delta n$ is the fluctuation in photon number and $\Delta 
\phi$ the error in measurement of a phase.  
Poisson statistics suggests $\Delta n \sim \sqrt{n}$ which we choose to 
be conservative.  In effect, measurement of a phase to order one needs 
one photon, which we will estimate as ten.  Meanwhile the number of 
photons detectable $n_{flux}$ decreases somewhat faster than $1/r^{2}$. 
When the number of photons detected falls below what is needed for a 
reliable phase we reach an upper limit on distance measurement.  If we 
conservatively choose sources separated by $\Delta \theta \sim arcsec$, 
take $\Delta x \sim 10^{4} km$, and optical $\lambda \sim 5 \times 
10^{-4} \, mm$, then $\phi^{parallax}  \sim 1, \, \phi^{0} \sim 
10^{8}$, which is intimidating. Assuming $\vec \Delta x$ is allowed to 
drift, it must remain steady to one part in $10^{8}$ during a single 
phase measurement.  If $\vec \Delta x$ rotates once per day ($10^{5}$ 
sec), each phase measurement needs $10^{4}$ photons per second.  For 
the number of photons we rescaled the calculations of Ref.\cite{ulmer}. 
  We assumed a 100 $m^{2}$ aperture with 10\% throughput efficiency, and 
  Hubble constant $H =0.7 \cdot100 km/s/Mpc$.  To order of magnitude 
  there are 20 photons per second in $H_{\a}$ light for a typical bright 
  galaxy with power $10^{42}$ ergs/s at redshift $z \sim1$: Insufficient 
  flux for a Gpc-range measurement.  However an AU-scale source at $100 
  \, Mpc$ range is about $10^{3}$ times more luminous, allowing the 
  measurement.  Moreover, approximately $10^{5}$ such measurements can be 
  repeated per day.   We can see no barriers of principle.

In summary, we have concentrated on ambitious measurements with amplitude and
intensity correlations at the $Gpc$ scale because they seem to us the ultimate ambition.  From near Earth-orbit, or perhaps with detectors fixed on Earth, it should be possible to measure to $r \geq Mpc$, which would be magnificent.  A number of radio telescopes in orbit can make mutual correlations, and correlations with ground-based receivers, to develop stupendous resolution. The joint Japanese/US $VSOP$ (VLBI Space Observatory Program) mission had a 21000 $km$ orbit and an 8m telescope\cite{vsop}.  An ambitious Russian program RADIOASTRON plans for very large orbits\cite{radioastron}.  As far as we know both intensity and amplitude correlations can be performed with such instruments.  Meanwhile numerous topics in conventional astronomy would benefit enormously from direct measurements of distances thousands of times smaller than what we discuss, and correspondingly easier to implement.  While there are a host of challenging technical issues, there is every reason to believe that a new method of distance measurements should be feasible with current technology.

{\it Acknowledgments:} PJ thanks Vasant Kulkarni, Rajaram Nityananda
and Anantha Ramakrishna for useful discussions.  JP thanks Steve Myers informing us of VSOP and RADIOASTRON.  Research supported in part under DOE Grant Number
DE-FG02-04ER14308.

\end{document}